# Tracing molecular dynamics at the femto-/atto-second boundary through extreme-ultraviolet pump-probe spectroscopy

P. A. Carpeggiani[1,2], P. Tzallas[1*], A. Palacios[3], D. Gray[1], F. Martín[3,4] and D. Charalambidis[1,2]

[1]*Foundation for Research and Technology—Hellas, Institute of Electronic Structure and Laser, PO Box 1527, GR-711 10 Heraklion, Crete, Greece,*
[2]*Department of Physics, University of Crete, PO Box 2208, GR71003 Heraklion, Crete, Greece,*
[3]*Departamento de Química, Módulo 13, Universidad Autónoma de Madrid, 28049 Madrid, Spain*
[4]*Instituto Madrileño de Estudios Avanzados en Nanociencia (IMDEA-Nanociencia), Cantoblanco, 28049 Madrid, Spain*

*e-mail: ptzallas@iesl.forth.gr

**Coherent light pulses of few to hundreds of femtoseconds (fs) duration have prolifically served the field of ultrafast phenomena. While fs pulses address mainly dynamics of nuclear motion in molecules or lattices in the gas, liquid or condensed matter phase, the advent of attosecond pulses has in recent years provided direct experimental access to ultrafast electron dynamics. However, there are processes involving nuclear motion in molecules and in particular coupled electronic and nuclear motion that possess few fs or even sub-fs dynamics. In the present work we have succeeded in addressing simultaneously vibrational and electronic dynamics in molecular Hydrogen. Utilizing a broadband extreme-ultraviolet (XUV) continuum the entire, Frank-Condon allowed spectrum of $H_2$ is coherently excited. Vibrational, electronic and ionization 1fs scale dynamics are subsequently tracked by means of XUV-pump-XUV-probe measurements. These reflect the intrinsic molecular behavior as the XUV probe pulse hardly distorts the molecular potential.**

Electronic excitations in molecules are commonly in the VUV/XUV spectral region. Until recently most of the XUV sources were lacking either sufficient pulse energy (high harmonic generation (HHG) sources) or ultrashort pulse duration (free electron lasers (FEL)), thus preventing access to XUV-pump-XUV-probe measurements in the 1 *fs* or attosecond (*asec*) temporal scale. Experimental efforts in this time scale have been restricted to IR-XUV pump-probe schemes [1-7], or *in-situ* electron-ion collision methods [8]. Systematic developments in high pulse energy HHG [9-13] and XUV super-continua [14-18] paved the way to time delay spectroscopic studies [19] and XUV-pump-XUV-probe experiments [20-21] that have lately demonstrated their first proof of principle application in the measurement of induced, ultrafast evolving atomic coherences in an atomic continuum [19-21]. Such measurements are free of interventions from unwanted channels, opened through quasi-resonant multi-IR-photon transitions. The present work, motivated by a recent theoretical study [22], demonstrates the first 1 *fs* scale XUV-pump-XUV-probe study of ultrafast dynamics in $H_2$.

The scheme under investigation is shown in Fig. 1. The inset shows the spectrum of the continuum radiation used. Absorption of one photon excites coherently all allowed electronic states of $H_2$, in each of which a superposition of those vibrational levels that are within the Franck-Condon region are populated. A small fraction of the bandwidth ionizes the molecule leaving it in the bound part of the $H_2^+$ ionic ground state. The dynamics of the excited electronic and vibrational wavepackets are probed through absorption of a second photon from the second temporally delayed XUV pulse. Absorption of the second photon brings the molecule above its ionization limit at excess energies that allow fragmentation of the ion. Simultaneous absorption of two photons is energetically not sufficient to reach the $^2\Sigma_g^+(2p\sigma_u)$ dissociative state of the ion, producing mainly zero kinetic energy protons through dissociation of the ionic ground state $X\,^2\Sigma_g^+(1s\sigma_g)$. Upon evolution of the molecular vibration, the molecule stretches to internuclear distances from which the $^2\Sigma_g^+(2p\sigma_u)$ state is accessible through the absorption of the second photon. Fragmentation through this repulsive state produces protons with non-zero kinetic energies that depend on the internuclear distance at the moment of the absorption. Thus, this dissociation channel opens at a given delay after the first pulse excitation. Consequently, the delay is a parameter that can switch this channel *on* and *off* and control the kinetic energy of the produced protons. While two photons may also be absorbed either solely from the first or the second pulse, for the pulses used here the $^2\Sigma_g^+(2p\sigma_u)$ state is accessible only through one-photon absorption from the first pulse followed by one-photon absorption from the second one, i.e only through the pump-probe sequence.

The experimental set up used has been described in a previous work [14, 16, 19, 20, 23-25]. Charged interaction products, i.e. $H_2^+$ ions and $H^+$ fragments, are detected through a time-of-flight mass spectrometer. While specific proton kinetic energies cannot be resolved in the mass spectrum, protons with zero kinetic energy can be distinguished from those with non-zero kinetic energy in the mass peak structure. Zero kinetic energy fragments contribute mainly to the centre of the ion mass peak; while non-zero kinetic energy fragments are present mainly at the tails of the peak (see Supplementary Information).

Observable two-XUV-photon absorption is verified through intensity dependence measurements of the ion yields. The results are shown Fig. 2. In log-log scale the slope of $H_2^+$ yield is 1 as expected for a linear process, while the slope of the proton yield is close to 2 (1.7) indicative of a two-photon process.

It is worth mentioning that, as we will show below, the used XUV bandwidth is rather advantageous, as it is broad enough to efficiently launch a nuclear wave packet that favours sequential two-photon absorption to the $^2\Sigma_g^+(2p\sigma_u)$ continuum and decreases the relative importance of one- and two-photon absorption to the $X\,^2\Sigma_g^+(1s\sigma_g)$ continuum [26].

Fig. 3 depicts a measured temporal trace of the total proton yield as a function of the delay between the two XUV-pulses, while Fig. 4 shows the yield of protons with non-zero kinetic energy.

A non-typical but common feature in both figures is the local minimum observed at zero delay, i.e. at complete temporal overlap of the two XUV pulses, for which the optical interference is expected to lead to a maximum yield in a second order process. This minimum and the subsequent build up of the proton yield during the first *fs* are attributed to the following dynamics. At zero delay, the dissociation channel through the $^2\Sigma_g^+(2p\sigma_u)$ state is still energetically closed, while after $\approx 1$ *fs* the excited molecule stretches such that absorption of the second photon from the intermediate state reaches the repulsive state $^2\Sigma_g^+(2p\sigma_u)$ leading to an enhancement of the proton yield. At larger delays the yield decreases due to the wavepacket delocalization. Beyond their common feature around the zero delay, figs 3 and 4 preserve a significant difference, which is the signal ratio between near-zero and longer delay times. In the trace of fig. 3 the signal around zero delay is significantly stronger compared to that at longer delay times, a feature that is missing in the trace of fig. 4. This difference is associated to the excitation channels participating in each trace. The signal in fig. 3, results from the convolution of the pump-probe channel and the "direct" 2-photon absorption channel, i.e. the cycle averaged second order optical interference of the two pulses, while in fig. 4, contribution from the pump-probe channel, which meanly leads to non-zero kinetic energy protons, is larger. Thus, the increased yield in fig. 3, results from the convolution of the $2^{nd}$ order intensity volume autocorrelation (IVAC) maximum [23] with the dynamic effect of the dissociation channel opening. The non-zero kinetic energy protons of fig. 4 start being produced at delays for which only the tails of the two pulses are overlapping, and thus the optical interference effects is significantly reduced. Moreover two-photon absorption from each pulse does not essentially participate in this dissociation path, reducing the number of available proton production channels. The blue curve of fig. 4 has been obtained from theoretical calculations similar to those reported in [22], in which molecules are aligned parallel to the XUV field polarization and thus only states of $\Sigma$ symmetries are considered. In these calculations, the contribution from protons with KER<0.4 eV has been removed. For delays at which the two XUV pulses overlap the theoretical data have been corrected for the reduced peak to background ratio of the $2^{nd}$ order volume autocorrelation as compared to that of a conventional one. A slight difference in the delay for reaching maximum proton yield between theory (1.5 fs) and experiment (1 ± 0.2 fs) is due to the higher photon energies used in the experiment leading to conditions, where the $^2\Sigma_g^+(2p\sigma_u)$ state is accessible at smaller delays. This is verified by calculations performed at higher photon energy presented in the supplementary material section, where the narrow spike feature appearing in the calculated trace at zero delay is also discussed.

The traces of figs. 3 and 4 at longer delays depict similar features of multi-frequency beating. Unlike in many existing studies, this beating is the result of combined electronic and vibrational wavepacket dynamics. The concurrent coherent excitation of electronic and vibrational states allows the simultaneous tracking of electronic and nuclear motion. The Fourier transform of the trace in fig. 3 is shown in the lower panel. The pronounced peak at 0.09 fs$^{-1}$ corresponds to the half the vibrational period of the $C^1\Pi_u$, $B''^1\Sigma_u^+$, $D^1\Pi_u$ (unresolved) states at the excitation energy interval of the experiment [27]. The small peak at 0.04 fs$^{-1}$ is compatible with half the vibrational period of the $B^1\Sigma_u^+$ state but due to the maximum delay of the measured trace (±22fs) such a measurement is marginal (for the state assignments see Supplementary Information). The peaks at 0.44fs$^{-1}$ and 0.67fs$^{-1}$ can be attributed to beating frequencies between electronic states. They namely match the frequency differences between the $B^1\Sigma_u^+$ and $B''^1\Sigma_u^+$ and between the $B^1\Sigma_u^+$ and $B'''^1\Sigma_u^+$ electronic states respectively, while frequency differences involving Π symmetry states also contribute to these peaks. Frequency differences between higher electronic states may contribute to the peak at 0.22 fs$^{-1}$. More detailed spectral features could be resolved by increasing the total delay length.

Although the duration of the XUV pulses has not been measured, an upper limit can be extracted from the smallest measured structure widths in the traces. These are ≈ 1.5fs in the double peak structure around zero delay. Given that they result from a convolution of the pulse width and the width related of the physical processes involved, the XUV pulse duration should be 0.6 < $\tau_{XUV}$ < 1.5fs, 0.6 fs being its Fourier transform limited value. The pulse duration is not measured because of the not stabilized carrier-envelope-phase of the driving field, which causes shot-to-shot fluctuations of the XUV waveform alternating from single-pulse to double-pulse structures [20, 21, 28].

In a nut shell, utilizing an intense coherent XUV super-continuum radiation, simultaneously electronic, vibrational and dissociative ionization dynamics in molecular hydrogen, evolving at the 1 fs temporal scale, have been addressed by means of an XUV-pump-XUV-probe experiment. The present work paves the way to studies of dynamics beyond the Born-Oppenheimer approximation, addressing the dependence of the electronic dynamics on the variation of the internuclear distance during the vibrational motion. This will be achieved at increased temporal resolution, through reduced pulse duration, while recording of long temporal traces will substantially improve resolution in the frequency domain.

**Acknowledgements**

This work is supported in part by the European Commission programs ATTOFEL, CRISP, Laserlab Europe, the European COST Action MPI1203-SKO and the Greek funding program NSRF. A. P. and F. M. acknowledge allocation of computer time by CCC-UAM and BSC Mare Nostrum, and financial support from the Advanced Grant of the European Research Council XCHEM 290853, the European grant MC-RG ATTOTREND, the European COST Action CM1204 (XLIC), the MICINN project No. FIS2010- 15127, and the ERA-Chemistry project PIM2010EEC-00751.

**Figure Captions**

**Figure 1. The two XUV-photon resonant ionization/fragmentation scheme of $H_2$.** The inset shows the XUV spectrum used in the experiment. This spectrum is selected, using an Indium transmission filter, from the radiation produced through non-linear frequency up conversion applying the IPG to the HOHG process. Single photon absorption excites all dipole allowed electronic/vibrational states of $H_2$ within the Franck-Condon region. The evolution of the excited electronic/vibrational wavepackets is traced by a delayed replica of the XUV pulse, detecting the yield of the protons produced through fragmentation of the molecular ion. The vertical lines in the spectrum of the inset indicate the states that are resonant with the specific interval of the spectrum. (A) refers to the opening of the ionization/dissociation channel through the $^2\Sigma_g^+(2p\sigma_u)$ repulsive potential. (B) and (C) refer to the vibrational dynamics in the $B^1\Sigma_u^+$ (B) and $C^1\Pi_u / B'^1\Sigma_u^+ / D^1\Pi_u$ (C) intermediate electronic states of the neutral molecule. For the assignment see also Supplementary Information.

**Figure 2. Ion yield dependence of $H_2^+$ and $H^+$ on the XUV intensity.** The measured slopes of 1 and 1.7 are indicative of a linear and a two photon process underlying the production of $H_2^+$ and $H^+$, respectively.

**Figure 3. Measured temporal trace of the total proton yield and its Fourier transform spectrum.** The observed minimum at zero delay and the succeeding build up of the proton yield during the first 1 *fs* of delay is attributed to the dynamics of the opening of the dissociation channel through the $^2\Sigma_g^+(2p\sigma_u)$ repulsive potential. A, B, C as in figure 1. In the Fourier transform spectrum at the bottom the peak 0.09 fs$^{-1}$ correspond to the times the excited vibrational wavepackets need to reach the outer turning points in the $C^1\Pi_u / B'^1\Sigma_u^+ / D^1\Pi_u$ (unresolved) intermediate electronic states, respectively. The small peak at 0.04 fs$^{-1}$ could be assigned to the $B^1\Sigma_u^+$ state but its small amplitude does not permit a firm assignment. The frequency peaks at 0.44 fs$^{-1}$, 0.67 fs$^{-1}$, match the beating frequencies between the $B^1\Sigma_u^+$ and $B'^1\Sigma_u^+$ and between the $B^1\Sigma_u^+$ and $B'''^1\Sigma_u^+$ electronic states, respectively. Contributions from frequency

differences between higher Σ and Π Rydberg states may also contribute to the frequency peak 0.22 fs$^{-1}$.

**Figure 4. Measured temporal trace of the yield of non-zero kinetic energy photons.** The minimum at zero delay and the yield build up within the first *fs* is also observable here as in figure 3. The overall higher observed proton yield than anywhere else observed around zero delay in the trace of fig 3 is missing here. This observed increased yield in figure 3 is the result of the convolution of the optical interference and the dynamic effect of the dissociation channel opening. The blue curve is from theoretical calculations, in which only intermediate states of *Σ* symmetries have been considered, i.e. for molecules aligned parallel to the polarization of the XUV field. The slightly shorter than in theory delay time (1 *fs*) that the maximum is reached in the experiment can be attributed to the shorter wavelengths used in the present work. The inset on the right depicts two ion mass peaks measured at 1.1 and 12.5 fs delays. To the $H_R^+$ edge contribute non-zero kinetic energy proton fragments released towards the entrance of the TOF spectrometer. To the $H_L^+$ edge contribute non-zero kinetic energy proton fragments released towards the repeller of the TOF spectrometer that are reflected towards the TOF entrance by the repelling voltage. To the central $H_C^+$ part of the peak contribute protons produced with zero velocity component parallel to the TOF axis. At 12.5 fs delay the production of non-zero kinetic energy protons is minimized.

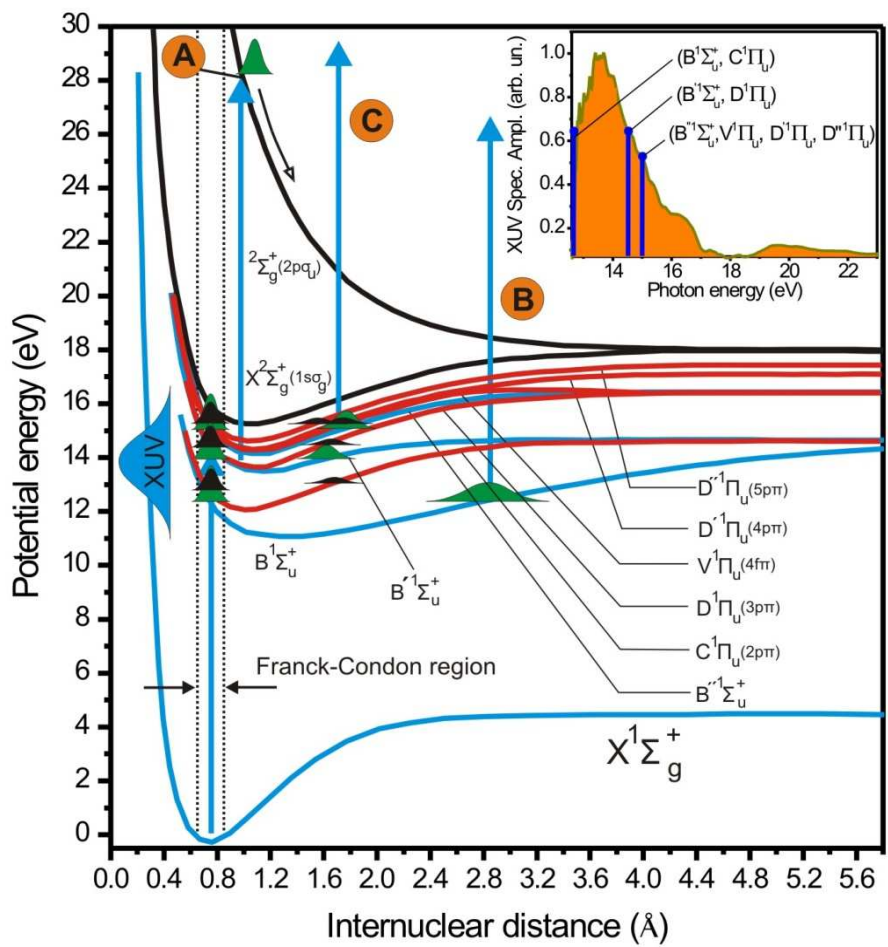

**Figure 1**

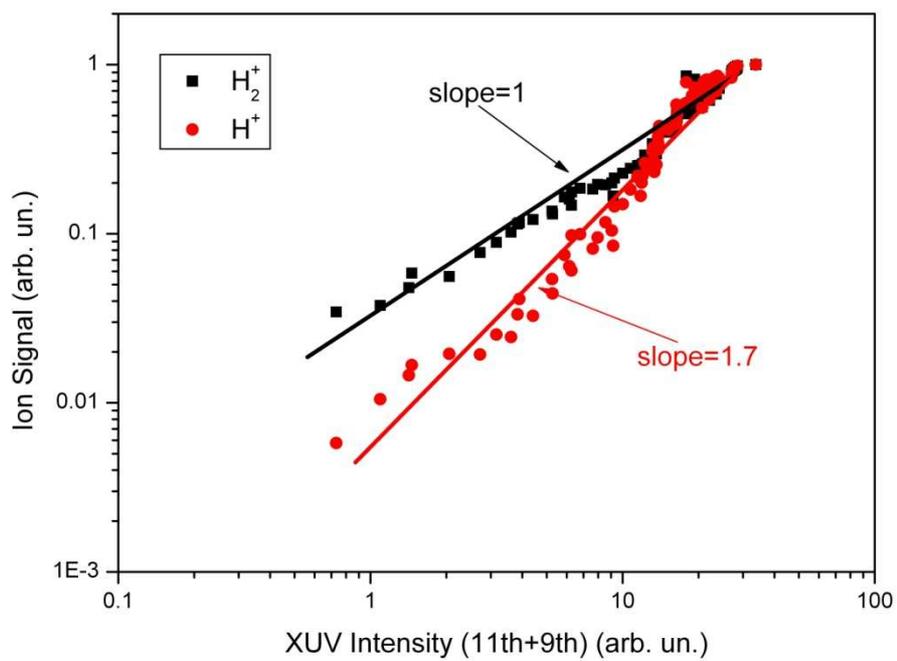

**Figure 2**

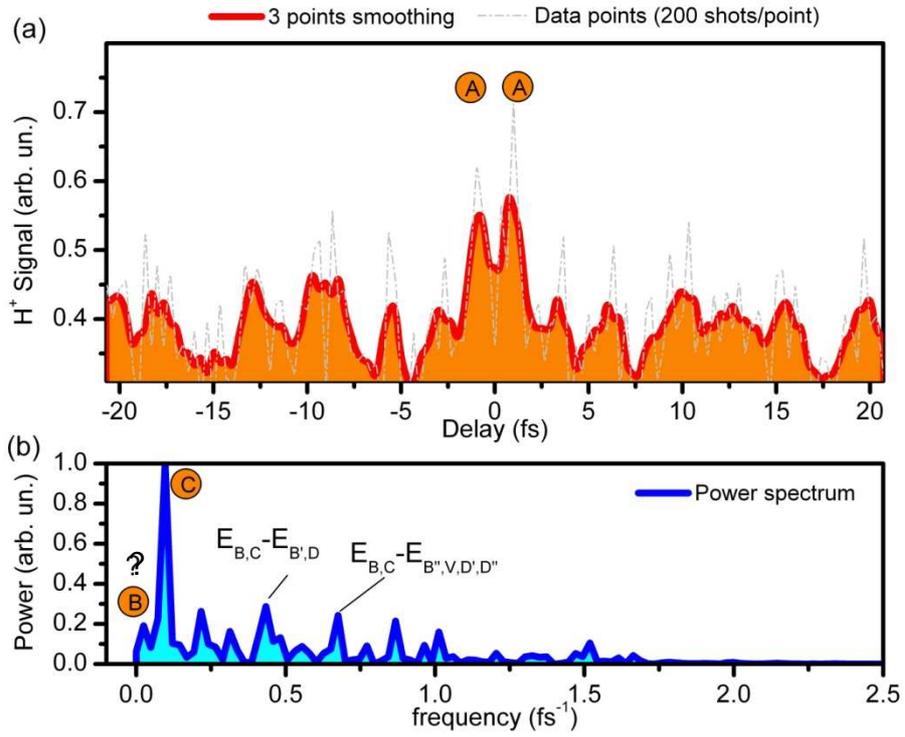

**Figure 3**

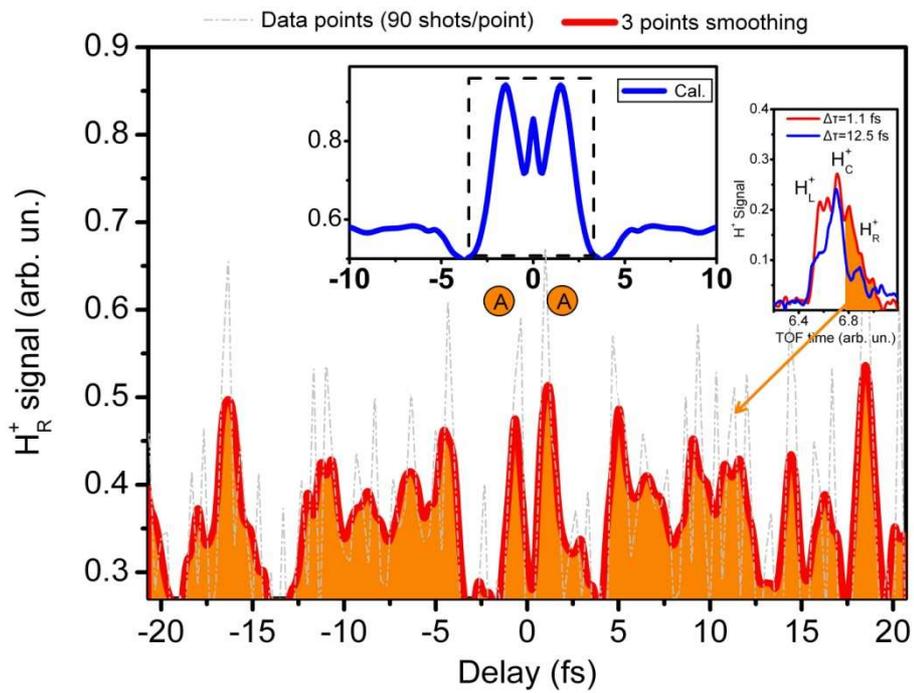

**Figure 4**